\documentclass[journal=jacsat,article]{achemso}

\usepackage[utf8]{inputenc} 
\usepackage[english]{babel}
\usepackage[T1]{fontenc}    
\usepackage{booktabs}       
\usepackage{amsfonts}       
\usepackage{caption}
\usepackage{subcaption}
\usepackage{hyperref}       
\usepackage{url}

\usepackage{pdfpages} 

\newcommand\MIC{\ensuremath{\mathit{MIC}}}
\newcommand\MICu{\ensuremath{\mathrm{\mu}\mathrm{g}/\mathrm{ml}}}

\usepackage[toc,page,titletoc]{appendix}
\usepackage{subfiles} 

\usepackage{titlesec}
\titleformat{\subsection}{\itshape\normalsize}{\thesubsection}{1em}{}[\vspace{-0.5ex}]

\title{Predicting antimicrobial activity of conjugated oligoelectrolyte molecules via machine learning}

\author{ \footnotesize Armi Tiihonen}
\email{armi.tiihonen@gmail.com \& tiihonen@mit.edu}
\affiliation[Massachusetts Institute of Technology]{ \footnotesize Massachusetts Institute of Technology, Cambridge, MA 02139, USA}

\author{Sarah J. Cox-Vazquez}
\affiliation[National University of Singapore]{Departments of Chemistry and Chemical \& Biomolecular Engineering, National University of Singapore, Singapore 119077, Singapore}
\email{sjcoxvazquez@nus.edu.sg}

\author{Qiaohao Liang}
\affiliation[Massachusetts Institute of Technology]{Massachusetts Institute of Technology, Cambridge, MA 02139, USA}

\author{Mohamed Ragab}
\affiliation[Nanyang Technological University]{School of Computer Science and Engineering, Nanyang Technological University, 50 Nanyang Avenue, Singapore 639798}

\author{Zekun Ren}
\affiliation[Singapore MIT Alliance for Research and Technology]{Singapore MIT Alliance for Research and Technology, \#05-09, Innovation Wing, 1 Create Way, Singapore 138602}

\author{Noor Titan Putri Hartono}
\affiliation[Massachusetts Institute of Technology]{Massachusetts Institute of Technology, Cambridge, MA 02139, USA}

\author{Zhe Liu}
\affiliation[Massachusetts Institute of Technology]{Massachusetts Institute of Technology, Cambridge, MA 02139, USA}
\altaffiliation{This author is now at: School of Materials Science and Engineering, Northwestern Polytechnical University (NPU), Xi'an, Shaanxi 710072 P. R. China}

\author{Shijing Sun}
\affiliation[Massachusetts Institute of Technology]{Massachusetts Institute of Technology, Cambridge, MA 02139, USA}

\author{Cheng Zhou}
\affiliation[National University of Singapore]{Departments of Chemistry and Chemical \& Biomolecular Engineering, National University of Singapore, Singapore 119077, Singapore}

\author{Nathan C. Incandela}
\affiliation[University of California]{Center for Polymers and Organic Solids, Department of Chemistry and Biochemistry, University of California, Santa Barbara, Santa Barbara, California 93106, USA}

\author{Jakkarin Limwongyut}
\affiliation[University of California]{Center for Polymers and Organic Solids, Department of Chemistry and Biochemistry, University of California, Santa Barbara, Santa Barbara, California 93106, USA}

\author{Alex S. Moreland}
\affiliation[University of California]{Center for Polymers and Organic Solids, Department of Chemistry and Biochemistry, University of California, Santa Barbara, Santa Barbara, California 93106, USA}

\author{Senthilnath Jayavelu}
\affiliation[Agency for Science, Technology and Research]{Institute for Infocomm Research, Artificial Intelligence, Analytics And Informatics, Agency for Science, Technology and Research, Singapore, 138632}

\author{Guillermo C. Bazan}
\affiliation[National University of Singapore]{Departments of Chemistry and Chemical \& Biomolecular Engineering, National University of Singapore, Singapore 119077, Singapore}
\alsoaffiliation[University of California]{Center for Polymers and Organic Solids, Department of Chemistry and Biochemistry, University of California, Santa Barbara, Santa Barbara, California 93106, USA}
\email{chmbgc@nus.edu.sg}

\author{Tonio Buonassisi}
\affiliation[Massachusetts Institute of Technology]{Massachusetts Institute of Technology, Cambridge, MA 02139, USA}
\alsoaffiliation[Singapore MIT Alliance for Research and Technology]{Singapore MIT Alliance for Research and Technology, \#05-09, Innovation Wing, 1 Create Way, Singapore 138602}
\email{buonassi@mit.edu}

\begin{document}

\clearpage

\begin{abstract}

New antibiotics are needed to battle growing antibiotic resistance, but the development process from hit, to lead, and ultimately to a useful drug, takes decades. Although progress in molecular property prediction using machine-learning methods has opened up new pathways for aiding the antibiotics development process, many existing solutions rely on large datasets and finding structural similarities to existing antibiotics. Challenges remain in modelling of unconventional antibiotics classes that are drawing increasing research attention. In response, we developed an antimicrobial activity prediction model for conjugated oligoelectrolyte molecules, a new class of antibiotics that lacks extensive prior structure-activity relationship studies. Our approach enables us to predict minimum inhibitory concentration for \textit{E. coli} K12, with 21 molecular descriptors selected by recursive elimination from a set of 5,305 descriptors. This predictive model achieves an $R^2$ of 0.65 with no prior knowledge of the underlying mechanism. We find the molecular representation optimum for the domain is the key to good predictions of antimicrobial activity. In the case of conjugated oligoelectrolytes, a representation reflecting the 3-dimensional shape of the molecules is most critical. Although it is demonstrated with a specific example of conjugated oligoelectrolytes, our proposed approach for creating the predictive model can be readily adapted to other novel antibiotic candidate domains.

\end{abstract}

\section{Introduction}

Antibiotic resistance is on the rise globally and confronts us with the potential of up to 10 million deaths per year by 2050 if no actions are taken \cite{WHO2019}. Even though new antibiotic candidates are in pre-clinical development pipelines \cite{Theuretzbacher2020, Coates2011NovelSame}, the development cycle remains slow, typically taking 10-15 years. Moreover, microbes inevitably build resistance to new chemical structures, thereby requiring multiple new classes of antibiotics to be continuously discovered. The successful development of truly new antibiotics will likely be challenged by minimal mechanistic insight on the candidates and few structural analogs available. One may invest in the detailed bio-centric traditional approach that identifies the point of action. Alternatively, a series of structural variations in the molecules are generated with the anticipation of drawing an empirical structure-activity relationship. Within the context of the latter, chemical intuition and inference are brought together to develop a more successful structure. Such is the situation with conjugated oligoelectrolyte (COE) molecules, a class of novel antibiotics candidates. In the case of COEs, the challenge is further exacerbated by their complex structure, absence of analogs, and gram-type specificity, to name a few. Machine learning (ML) may offer an alternative approach to streamline development, whereby the essential elements in the molecular structure are identified that most strongly correlate with antibiotic activity. However, thousands of descriptors may be generated for each COE molecule, thereby necessitating the development of a principled downselection process in order to develop an operational model. We solve this problem and propose a ML model for predicting antimicrobial activity of COEs.

COEs are a distinctive class of molecules unified by their hydrophobic $\pi$-electron conjugated core and pendant groups bearing ionic functionalities \cite{Zhou2019AStability, Wang2017AActivity} --- see Figure \ref{fig:coe} for the examples used in this work. The tunable and molecular COE framework offers a practical advantage for developing a new class of antibiotics as the structures can easily be tailored to a desired physical property \cite{Limwongyut2020MolecularCells, Zhou2018InformedActivity, Yan2016InfluenceOligoelectrolytes}. A common property of these compounds is their ability to intercalate and disrupt biological membranes, which is their presumed antibiotic mechanism of action \cite{Hinks2014ModelingTransfer}. This proposed mechanism is advantageous for developing new antibiotics, as bacteria are expected to be less likely to develop resistance to membrane disruptors, compared to specific receptors. However, there is an absence of a known specific binding site that may guide the design of chemical structures.

ML modelling of molecular properties is enabled by recent developments in applied ML, especially regarding deep learning and molecular representations as previously reviewed elsewhere \cite{Walters2021ApplicationsPrediction, Elton2019, Chuang2020, Swann2018RepresentingLearning, David2020, Chuang2020}. Molecular representations include string representations such as widely used SMILES (Simplified Molecular Input Line Entry System) \cite{Weininger1988SMILESRules} or SELFIES (Self-Referencing Embedded Strings) designed to be used with ML methods \cite{Krenn2020Self-referencingRepresentation}. Numeric vectors consisting of molecular descriptor values (also called features) have been utilized already before the era of ML in QSAR (Quantitative Structure-Activity Relationship) modelling \cite{Danishuddin2016}, and have been formulated traditionally by feature engineering relying on domain knowledge, or by attempting to form general fingerprints such as Morgan fingerprints \cite{Morgan1965TheService}. Successful fingerprinting is a challenging task \cite{Capecchi2020OneMetabolome}, which is one of the reasons why molecular graphs networks and other learned representations have started to gain popularity over fingerprints. Molecular graph networks usually describe molecules with atoms as nodes, and with bonds between the atoms as edges. Graph networks have resulted in comparative accuracy with less samples in some benchmarking tests \cite{Wu2018MoleculeNet:Learning}, but there are also opposing results \cite{Jiang2021CouldModels}. ML model and molecular representation approaches can be closely tied together. \textit{E.g.}, message passing neural networks are combinations of a molecular graph network and a neural network, in which information is shared between the near neighboring atoms by passing messages between them \cite{Yang2019AnalyzingPrediction, Axelrod2021, Wu2018MoleculeNet:Learning}. Tree-based models such as random forest regression (RF) or gradient boosting also remain as an effective approach. Progress in the model analysis also facilitates increasing use of applied ML methods. New ML model analysis methods alleviate the black box challenge of ML in cases when it is necessary to use models that are not directly interpretable. Shapley additive explanations (SHAP) \cite{Lundberg2017APredictions, Lundberg2020FromTrees, Rodriguez-Perez2020InterpretationValues, Hartono2020a} is one of such approaches utilized in this work. It is a game-theoretic approach that differs from traditional feature importance ranking within models by connecting optimal credit allocation from a model input feature with local explanations of the model.

These developments in molecular representations and ML approaches are beginning to impact antibiotics discovery. The progress has already lead into the identification of a new antibiotic molecule: halicin had existed previously but was not known to act as an antibiotic before Stokes \textit{et al.} identified it by using solely a combination of deep learning --- directed message passing neural networks that are a combination of a feed-forward neural network and a molecule graph with directed message passing  scheme  \cite{Yang2019AnalyzingPrediction} --- and screening of large databases of thousands of existing molecules. Many of the approaches in ML-aided antibiotic property prediction have relied on large datasets and focus on finding new antibiotics candidates that resemble existing ones (\textit{i.e.}, have identical distributions). Such a setting is not readily available when considering new mechanisms of action or new molecular types as described above, while these approaches are gaining increasing attention in the preclinical development \cite{Theuretzbacher2020}.

Herein, we show a ML approach that may be applied to the challenge represented by new antibiotic classes, such as COEs. We present a framework for quickly establishing a predictive model of an antibiotic property. It consists of four components: (1) molecular representation, (2) feature down-selection, (3) ML algorithm selection, and (4) molecular descriptor importance analysis. We apply this framework to conjugated oligoelectrolyte molecules (COEs), down-selecting from 5,305 features to 21 critical features governing antimicrobial activity. With only 136 compounds measured, we demonstrate antimicrobial activity prediction with an accuracy of R$^2$ = 0.65. This framework does not rely on prior domain knowledge and is therefore compatible with novel molecule domains. The trained model, together with molecular descriptor -importance analysis and domain expertise, could serve as a foundation for the accelerated development of novel COEs with enhanced antimicrobial activity.

\section{Results \& Discussion}

\subsection*{Synthesis of 136 conjugated oligoelectrolyte molecules}

An experimental dataset covering 136 COE structures (Figure \ref{fig:coe}) was collected for this work. 113 of the molecules are newly synthesized and presented for the first time in this work. The remaining molecules have been presented before \cite{Zhou2018InformedActivity, Yan2016InfluenceOligoelectrolytes, Limwongyut2020MolecularCells, Zhou2019AStability, Hinks2014ModelingTransfer, Yan2015MembraneIntercalatingSystems} and are designated in the Supplementary Material. The COEs were characterized by NMR or mass spectrometry (shown in Supplementary Section S9) to confirm the synthesized structures. The molecules were screened for antimicrobial activity against \textit{E. coli} K12 bacteria by measuring minimum inhibitory concentration (\MIC{}) for bacteria growth. The COEs are subdivided into 9 categories based on the structure of the core and the number/composition of the pendant groups.
\begin{figure}[H]
  \caption{Molecular structures of COEs utilized in this work. Molecules are sorted into classes based on the core structure and ordered based on increasing length and complexity of the side chains. Counter ions not shown. (1) disubstituted stilbenzene COEs (2) di substituted azobenzene COEs (3) disubstituted styrylstilbenzene COEs (4) trisubstituted stilbenzene COEs (5) tetrasubstituted stilbenzene COEs (6) tetrasubstituted styrylstilbenzene with ethyl/propylamine-based side chains COEs (7) tetrasubstituted styrylstilbenzene with butylamine-based side chain COEs (8) Hexasubstituted styrylstilbenzene COEs  and (9) Miscellaneous COEs which did not fall into the other categories. Experimentally determined MIC values against against \textit{E. coli} K12 are stated under each molecule (ND = not determined for that COE).}
  \label{fig:coe}
\end{figure}

\begin{figure}[H]
    \addtocounter{figure}{-1} 
    \begin{subfigure}{\textwidth}
        \caption{}
        \centering
        \includegraphics[page=1, scale=0.285]{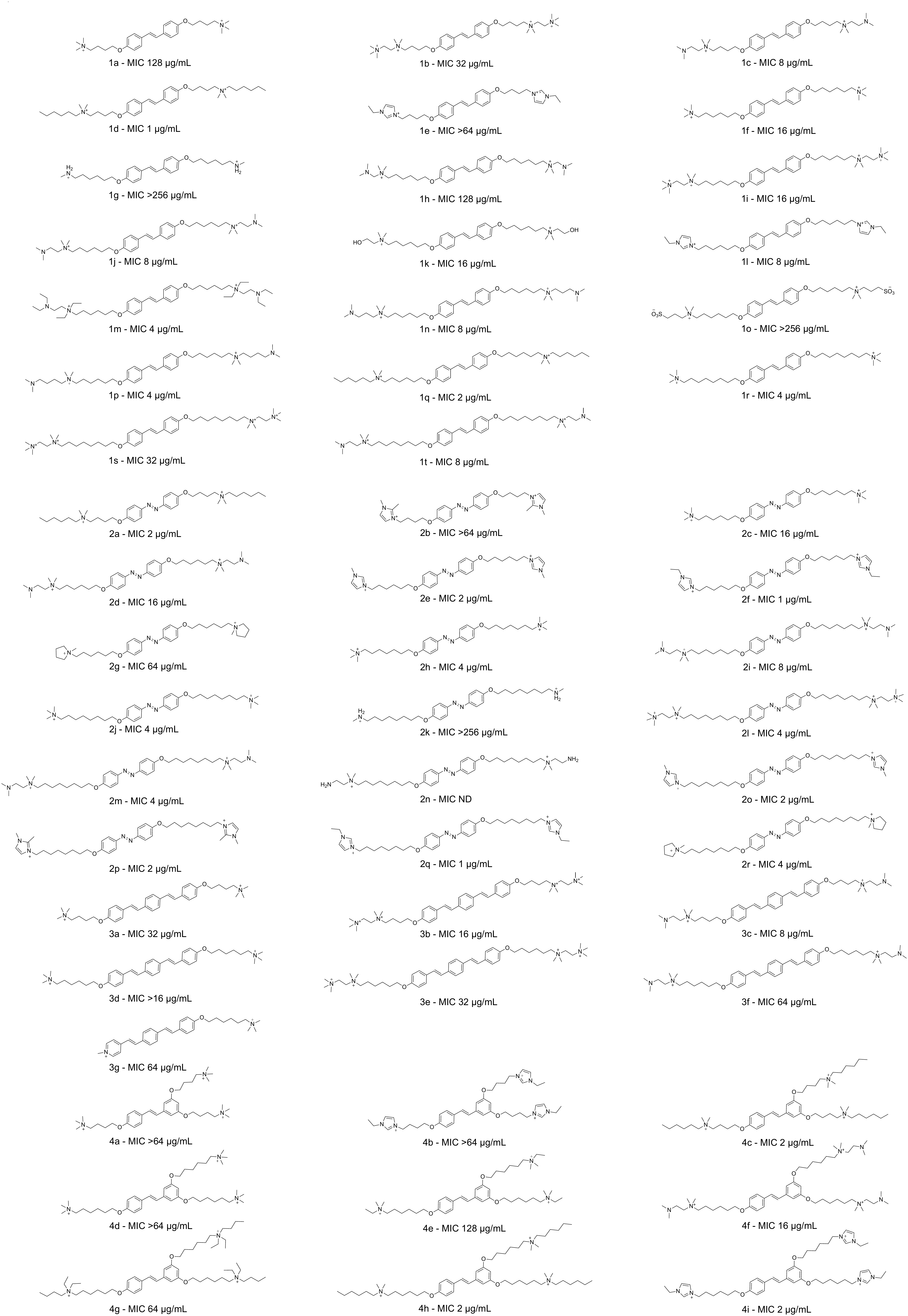}
    \end{subfigure}
\end{figure}

\begin{figure}[H]
    \ContinuedFloat 
    \begin{subfigure}{\textwidth}
        \caption{}
        \centering
        \includegraphics[page=1, scale=0.285]{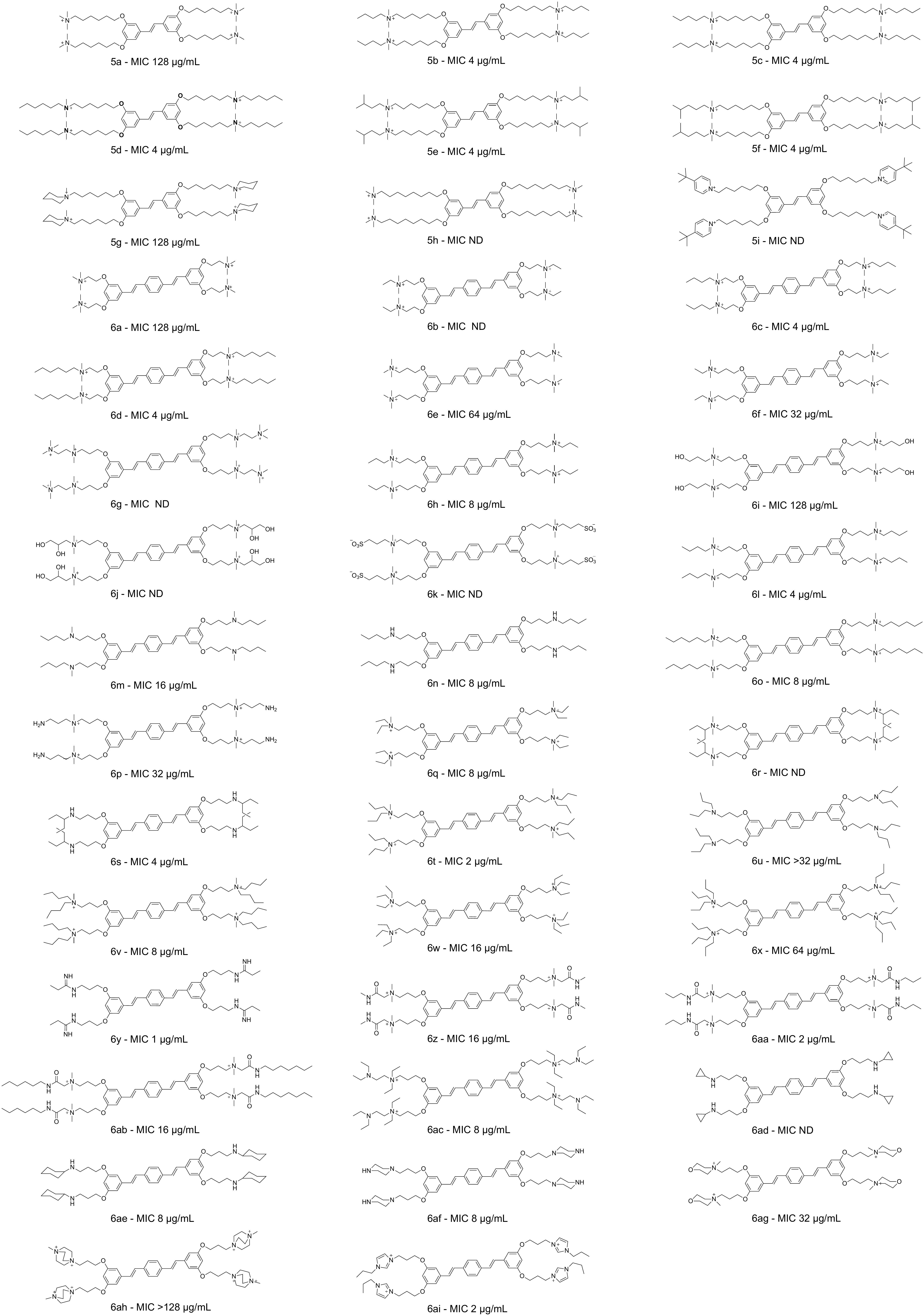}
    \end{subfigure}
\end{figure}

\begin{figure}[H]
    \ContinuedFloat 
    \begin{subfigure}{\textwidth}
        \caption{}
        \centering
        \includegraphics[page=1, scale=0.285]{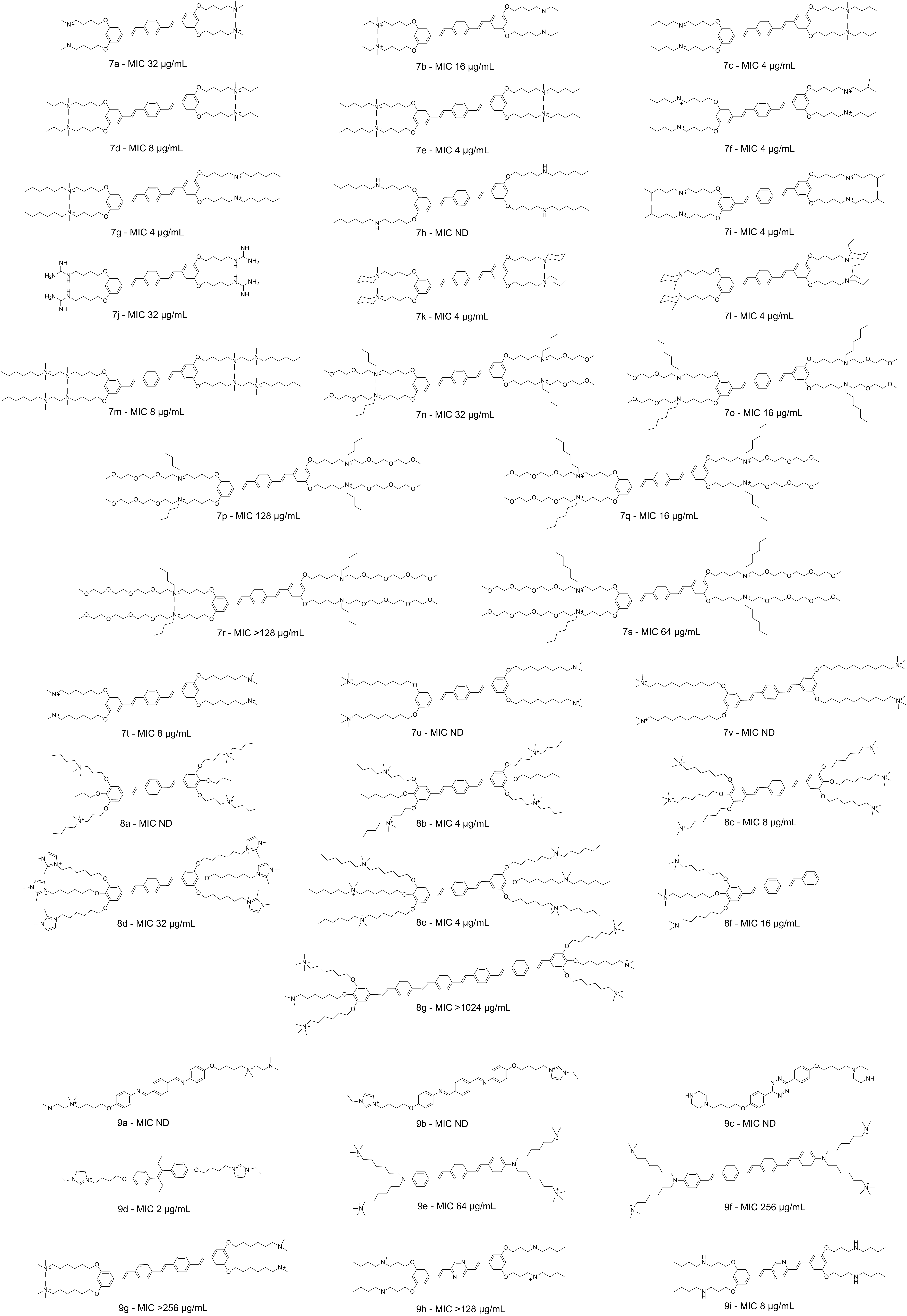}
    \end{subfigure}
\end{figure}

\subsection*{Principled downselection process}

\begin{figure}[h!]
    \centering
    \includegraphics[width=0.99\textwidth]{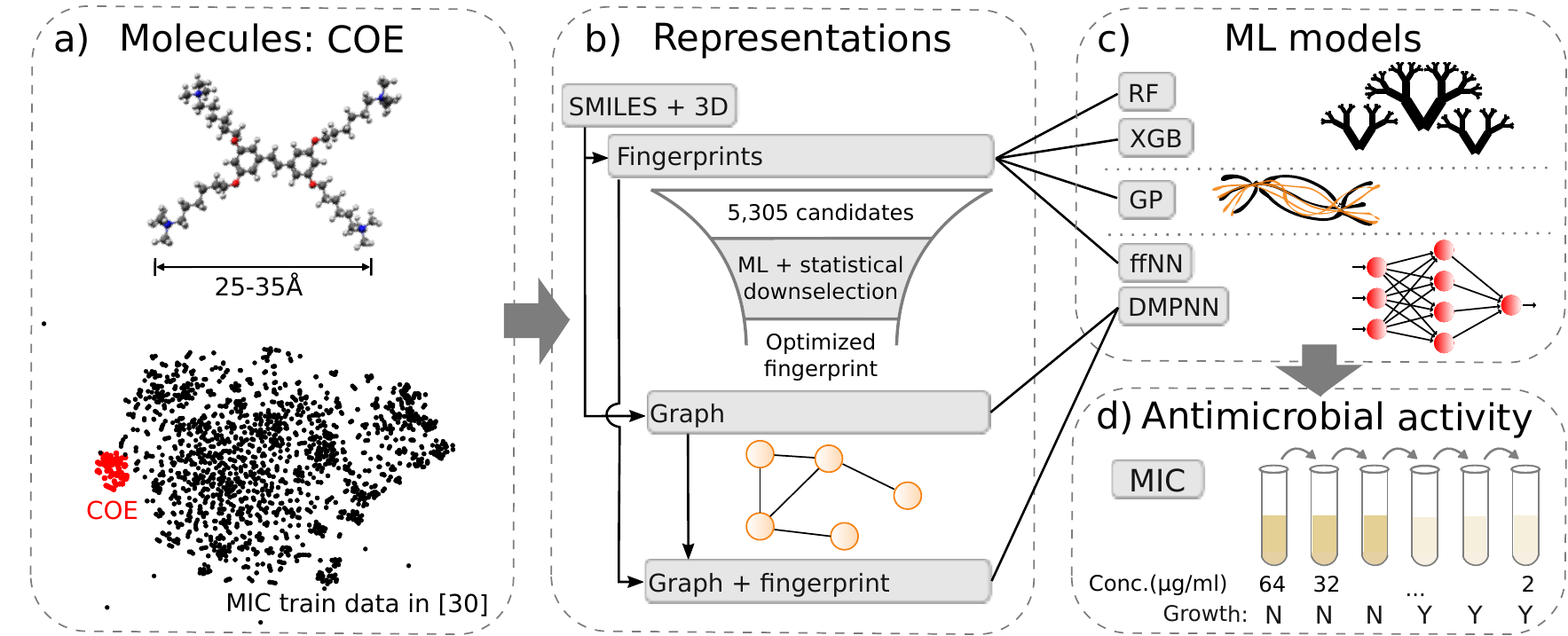}
    \caption{Predicting antimicrobial activity of conjugated oligoelectrolyte molecules (COE) involves optimizing molecular representation for COE molecules and pairing that with a suitable machine learning (ML) model. a) Example COE molecule (5a) and a similarity plot (t-Distributed Stochastic Neighbor Embedding with Morgan fingerprint and Jaccard similarity) with COEs (red) and an approximately 2,000-molecule dataset from Stokes \textit{et al.} \cite{Stokes2020} (black) that they utilized for training a model to predict molecules with antimicrobial activity. b) The molecular representation types investigated in this work. Molecular fingerprint candidates include a fingerprint optimized for COE by downselection from a base set of 5,305 candidate descriptors. c) ML models investigated in this work (definitions in the main text). d) Minimum inhibitory concentration (\MIC{}) for bacteria growth can be measured by diluting solutions until they cease to prevent \textit{E. coli} K12 bacteria growth. Two‐fold dilution down to 1 \MICu{} is one of the widely used approaches. The ML models are traied to predict 2-base logarithm of \MIC{}.}
    \label{fig:1}
\end{figure}

Our process for predicting the antimicrobial activity of COE molecules is illustrated in Figure \ref{fig:1}. Figure \ref{fig:1}a shows a similarity plot (t-Distributed Stochastic Neighbor Embedding with Morgan fingerprint and Jaccard similarity \cite{VanDerMaaten2008}) involving COEs and reference molecules, that are approximately 2,000 antibiotic candidate molecules previously tested for antimicrobial activity and used as a training dataset by Stokes \textit{et al.} \cite{Stokes2020}. The dataset was chosen as a reference due to its large size and diversity within the set. Analysis shows that COE molecules form a structurally distinct group of molecules compared to many other molecules that have previously been screened for antimicrobial activity, which warrants the development of a predictive ML model designed specifically for COE. One of the differences is size. Most existing antibiotics and many reference molecules included into the similarity plot of Figure \ref{fig:1}a are small compared to COEs (that are illustrated in Figure \ref{fig:coe}). 

Firstly, a molecular representation that is capable of capturing, and ideally even simplifying, the complexity of the structural and/or chemical factors driving the molecular activity is required. Secondly, this representation needs to be paired with a matching ML model. Here, we compare multiple molecular representation options: molecular fingerprints optimized for COE by downselecting descriptors as well as existing fingerprints from literature (described later), molecular graphs, and combinations of both (Figure \ref{fig:1}b). In this work, we define a fingerprint as a molecular descriptor vector that can contain any numeric descriptors in addition to binary ones. The representations are paired with decision tree -based ML models, a kernel method, and neural networks (Figure \ref{fig:1}c) to cover a range of widely used and advanced ML model options. We train the models to predict antimicrobial activity against \textit{E. coli} K12 bacteria, which has been measured experimentally as \MIC{} for bacteria growth (Figure \ref{fig:1}d). The ML models are trained to predict antimicrobial activity in the form of $\log_2(\frac{\MIC{}}{1\MICu{}})$ due to the base-2 exponential nature of \MIC{} values. The lower the model output value is, the higher is the predicted antimicrobial activity. 

We applied data curation and dropped potential outliers (5 molecules) from the data as pre-treatment steps (described in Methods and Supplementary Materials Section S13.1). These steps are crucial when treating experimental datasets, especially small datasets, because few outliers might distort the model significantly. The molecules with experimentally measured \MIC{} were split into 80\%-20\% train/test datasets. Throughout the results section, the performance of models and fingerprints are compared based on cross-validation with 20 stratified subsampling repeats of the training dataset (described in Section \ref{sec:methods}).

\subsection*{Optimized molecular fingerprint for conjugated oligoelectrolytes}

\begin{figure}[h!]
    \centering
    \includegraphics[width=0.85\textwidth]{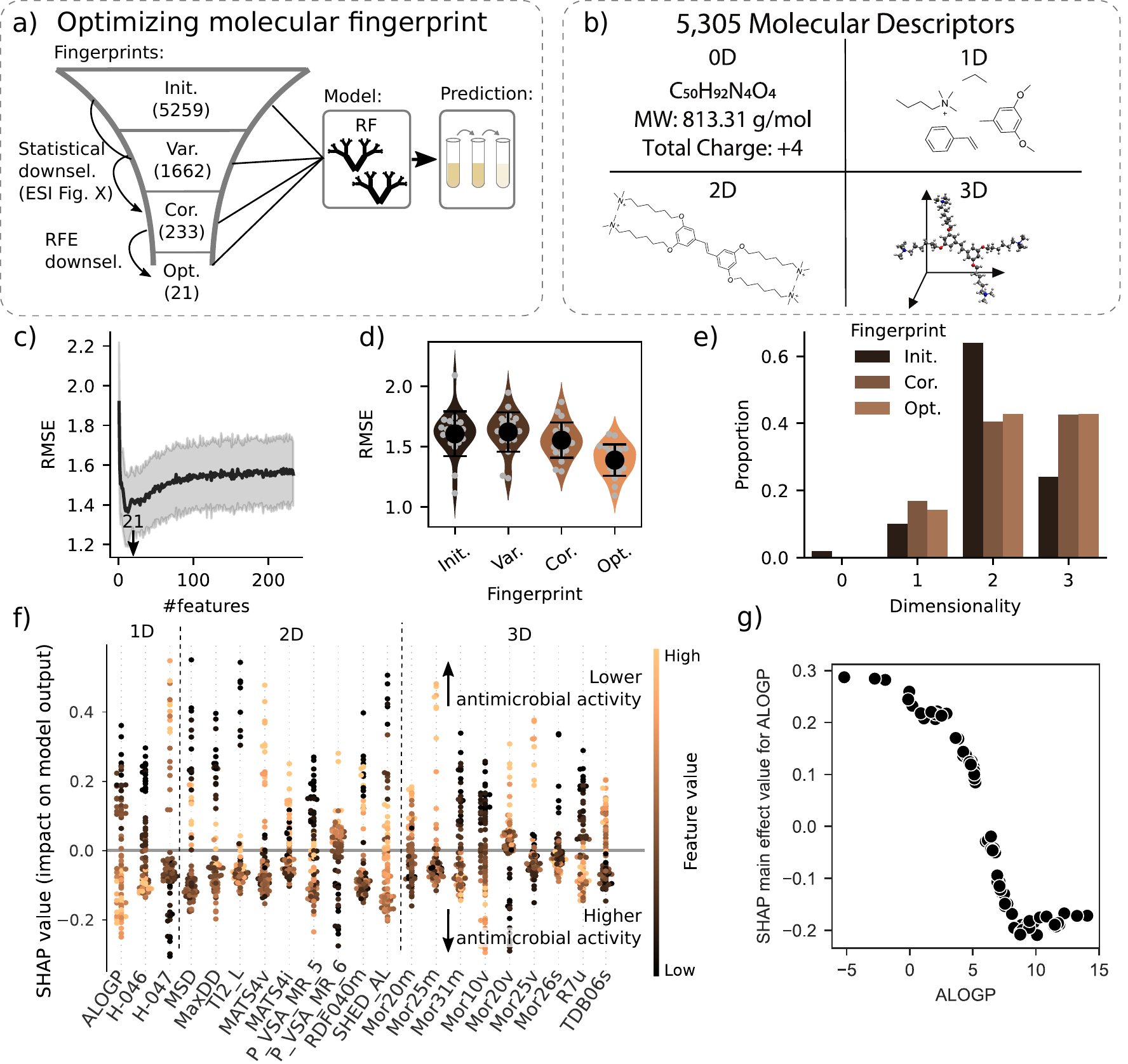}
    \caption{Optimization of molecular fingerprint for COEs improves the accuracy of the random forest regression (RF) model in predicting antimicrobial activity. a) Downselection process from a candidate descriptor set involving two stages of statistical downselection and one stage of ML downselection via recursive feature elimination (RFE). Fingerprints are defined in the main text and the lengths are shown in parentheses. b) Init. dataset included descriptors related to zero- to three-dimensional shape of the molecules (explanations in the main text). c) Root-mean-square error (RMSE; mean and standard deviation, std., are shown) during RFE shows a minimum error region centred at 21 descriptors, which are chosen as the Opt. fingerprint. RMSE is from cross-validation with 20 stratified subsampling repeats of the training dataset. d) RMSE for RF models trained with each fingerprint. Violins represent the distributions of the values; individual subsamples are shown in gray, and mean and std. of RMSE in black. e) Proportion of descriptors related to the 1D-3D properties of the molecules in the specified fingerprints. f) Shapley additive explanations (SHAP) analysis of the training dataset molecules and the final RF model (trained with the whole training dataset and Opt. fingerprint). Negative SHAP values push the model prediction towards low MIC (high antimicrobial activity), and vice versa. AlvaDesc naming convention is used for the descriptors. g) SHAP value for Ghose-Crippen octanol-water partition coefficient (ALOGP) of the training set molecules as a function of descriptor value.}
\label{fig:2}
\end{figure}

We began by optimizing a molecular fingerprint for COE. We aimed to avoid biasing our fingerprint toward any potential mechanism of antimicrobial activity, by not using expert knowledge in the fingerprint-selection process. Instead, the process consists of multiple stages of automated molecular-descriptor downselection, starting from a broad set of candidate descriptors (see Methods for detailed descriptions of the steps and Figure \ref{fig:2}a for process illustration). We initially generated 5,305 molecular descriptor candidates with the chemistry analysis software AlvaDesc from the SMILES strings and 3D structures of each COE. The set is a diverse combination of descriptors related to molecule chemistry and structure (full list shown in \cite{AlvaDescAlvascience}, more information in \cite{Todeschini2000HandbookDescriptors}). Descriptors can be classified based on their dimensionality when considering how they approximate the three-dimensional (3D) shape of a molecule (Figure \ref{fig:2}b). Our candidate descriptors include 0D (with no relation to shape, \textit{e.g.}, molecular weight), 1D (\textit{e.g.}, presence of certain active substructures within the molecule), 2D (\textit{e.g.}, molecular graph representations involving bonds between atoms but not bond lengths), and 3D (\textit{e.g.}, distances between certain atomic pairs in the molecule) ones. This broad set of descriptors is likely to contain information relevant for the antimicrobial action of a given COE. However, ML models are prone to overfitting with a large number of input parameters and small training data. This limits the options of viable ML models and is also likely to reduce the predictive accuracy of those options that remain to be viable. Therefore, we aimed for removing redundant and irrelevant descriptors that the 5,305-descriptor dataset would be anticipated to contain.

Of the 5,305 total descriptors, 5,259 are numerical descriptors, which we refer to as initial fingerprint (Init.). We applied two stages of statistical downselection and one stage of ML-aided descriptor downselection (Figure \ref{fig:2}a): Constant or almost constant descriptors were filtered out (variance downselection with relative limit of 0.1, Var. fingerprint) using the whole COE dataset. Next, the descriptors with high Spearman rank correlations between each other or a low correlation with \MIC{} were dropped (correlation downselection with limits of 0.9 and 0.05, respectively, Cor. fingerprint). Conservative variance and correlation limits were chosen for statistical downselection. Nevertheless, they allow for drastic reduction in the number of descriptor candidates (from 5,259 to 1,662, and then to 233), while restricting the likelihood of losing relevant descriptors (the resulting correlation matrices are shown in Supplementary Figure S1). The final, ML-aided stage of downselection was recursive feature elimination (RFE) \cite{Guyon2002GeneMachines}, which eliminates features based on feature importance rankings of the random forest regression (RF). A same kind of approach could have been adopted also with other ML model types, however, RF is a good option due to its straightforward training and feature importances. Wrapper methods, such as RFE, ensure that the resulting fingerprint carries the most important information by actually training ML models. RFE should complement the statistical filtering stages that ensure the breadth of information among the descriptor candidates. RFE was applied on Cor. fingerprint and training dataset in a mode that drops one descriptor at each recursion, see Methods for detailed information. Root-mean-square error (RMSE) graph in Figure \ref{fig:2}c shows that the optimum predictive accuracy of RF is reached in a region centered to fingerprint length of 21 descriptors. These descriptors were consequently chosen as the final fingerprint optimized for COEs (Opt., see Supplementary Material Section S2 for descriptors).

Next, RF models trained with the downselection fingerprints are investigated for the purposes of comparing the fingerprints. The cross-validation RMSE of the resulting RF models decreases along the downselection stages (Figure \ref{fig:2}d), showing that the downselection process improves the fingerprints. Opt. fingerprint provides average RMSE improved by 20\% in comparison to the Init. fingerprint (additionally, the test dataset predictions from the RF models trained with the full training dataset are shown in Supplementary Figure S2). While the COE dataset may be extensive from a synthetic viewpoint, it remains rather limited in size compared to typical datasets for ML. Thus, the consistency of descriptor downselection results needs to be evaluated carefully. We repeated the RFE with another tree-based ML model, gradient boosting regression (XGB). It resulted in very similar downselected fingerprint (Supplementary material Section S4), thus suggesting that the results are not an arbitrary result of the chosen surrogate ML model.

\clearpage

\subsection*{Analysis of molecular fingerprint optimized for conjugated oligoelectrolytes}

The molecular fingerprint optimization process is useful not only because it optimizes predictive accuracy of the resulting ML model, but also because the fingerprint content provides clues for understanding the underlying mechanisms of action for the molecules. Figure \ref{fig:2}e illustrates that Opt. fingerprint consists of 1D–3D descriptors with a large proportion of them being 3D. Cor. fingerprint, that also links directly to \MIC{} via statistical correlations, has very similar proportions. A large proportion of 3D descriptors compared to the Init. fingerprint survives until the last stages of downselection, which suggests that the antimicrobial activity of COEs is influenced by 3D molecular shape. This supports the presumed antibiotic mechanism of action for COE, that is the disruption of microbial membranes \cite{Hinks2014ModelingTransfer}. In our work, 3D shapes of COEs have been determined by molecular dynamics simulations with a relatively fast method that can contain small differences to the actual shape of the molecule. This fact, together with the dataset size, limits the extent of conclusions we can draw in this study based on individual 3D descriptors, but some observations seem consistent when analyzing the whole fingerprint.

Eight descriptors, more than one third of the Opt. fingerprint, are 3D-MoRSE descriptors (Molecular Representation of Structures based on Electronic diffraction \cite{Schuur1996TheActivity, Devinyak2014, Saiz-Urra2006QSARSelection}). These descriptors sum up contributions from each atom of the molecule based on pairwise distances of the atom pairs. Here, MoRSE descriptors with scattering factors 31, 26, 25, 20, and 10Å$\mathrm{^{-1}}$ survived the downselection, and are weighed by atomic mass (m), van der Waals volume (v), or intrinsic I-state (s). The high-order scattering factors suggest that subtle changes in the distances of atomic pairs (\textit{e.g.}, $2\pi/20$Å$^{-1}\approx0.3$Å) and short-range interactions overall might affect the antimicrobial activity of COE. The rest of the 3D descriptors in the Opt. fingerprint are a GETAWAY descriptor and a topological distance descriptor both related to autocorrelations \cite{Consonni2002Structure/responseDescriptors}.

From molecular property prediction viewpoint, the large proportion of 3D descriptors is notable because 3D molecular representations are not commonly used in antimicrobial activity prediction. Additionally, most of the existing such works focus on peptides \cite{Liu2018NovelPeptides, Khalil2020SynthesisComplexes} instead of small molecules. There are several reasons for the present scarcity of the use of 3D representations. First, many of the existing known antibiotic molecules are small and may operate via mechanisms of action that are dependent on intermolecular interactions, \textit{e.g.} \textit{via} targeting certain enzymes \cite{Boehm2000NovelScreening, DePascale2010AntibioticSolutions, Guo2011ASuperbug}. Thus, 3D representations may not have as strong an effect when searching for molecules with this type of operation, since they relate more to intramolecular interactions. The use of 3D molecular representations may not be necessary in cases when molecules with similar operation to the predecessors are searched for. Second, the benefits of 3D representations in molecular property prediction generally are currently not fully clear. For example, 3D molecular graph networks have resulted in a similar or even decreased accuracy with many prediction tasks focusing on small molecules \cite{Wu2018MoleculeNet:Learning, Axelrod2021} but nevertheless improved predictions of some, \textit{e.g.} quantum mechanical molecule properties \cite{Wu2018MoleculeNet:Learning} derived directly from the physical properties of the molecules. Performance has also depended on the exact model-representation combination \cite{Axelrod2021}. Third, generating 3D structures of molecules requires additional effort and computation time in comparison to representations that can be generated solely from \textit{e.g.} SMILES strings. Our results show that the effort is worthwhile in the case of COEs. This result may generalize into other antibiotic candidate domains with larger molecular structures. Incorporating the 3D shape of the molecules should therefore be considered in these cases.

To gain chemical insight into the relative contributions of different features, we investigated the descriptors in the Opt. fingerprint set with SHAP analysis. The method itemizes the contributions from each molecular descriptor to the predicted antimicrobial activity of each molecule when the ML model makes predictions. In Figure \ref{fig:2}f, SHAP is applied to analysing predictions of the training dataset molecules using the RF model trained with Opt. fingerprint (which also coincides with the best-performing model-representation pair for COEs as we show later). There are no clear dominating descriptors, but the high antimicrobial activity builds up from multiple contributions. The impacts from different descriptors are not reaching consistently high or low within the same molecule, either. Instead, for most COEs, some of the descriptors reduce the model's predicted MIC value, while some other descriptors increase it.
The complex feature interactions hinted by SHAP suggest why the antimicrobial activity of COE is challenging to predict. Similar reasons may explain the complexity of predicting antimicrobial activity in other antibiotic candidate domains.

We highlight the behavior of one of the Opt. descriptors, water-octanol partition coefficient (ALOGP), to suggest future direction for ML in antibiotics development. SHAP analysis shows that ALOGP should be high in order to reach high antimicrobial activity (Figure \ref{fig:2}g). The observation is, again, in line with the presumed mechanism of action for COE: high ALOGP indicates high lipophilicity of the molecule, which is a major factor determining how easily the molecules penetrate lipid bilayers. The ML model trained only for antimicrobial activity could not, however, recognize that the guideline of high lipophilicity is in practice bound by other antibiotic development criteria \cite{Lipinski2001ExperimentalSettings}, \textit{e.g.} cytotoxicity to human cells. This is because too high lipophilicity tends to lead to less targeted effects in the human body \cite{Arslan2020AApproach}. The case of ALOGP exemplifies that when investigated from the viewpoint of accelerating antibiotics development pipelines, ideal ML models do not only max out a single antibiotic development criterion but guide efforts towards candidates that fulfill several criteria adequately. Therefore continued efforts to develop reliable multi-task models \cite{Rusu2016, Sosnin2019AChemoinformatics, Ramsundar2017IsPharma, 2003DeterminationDilution, Camacho2018} are valuable even though they require more training data than what is available for COEs at the moment.

\subsection*{Comparison of molecular representation - machine learning model pairs}

While the preceding analysis focused on the optimized molecular fingerprint, and the RF model provided clues for factors contributing to COE activity, further comparisons of ML models and molecular representations are required to determine which combination has the highest predictive accuracy of antimicrobial activity for COEs. As illustrated in Figure \ref{fig:1}, ML models tested include tree-based ML models (RF and gradient boosting regression, XGB), a kernel method (Gaussian process regression, GP), and neural network models (feed-forward neural network, ffNN, and directed message passing neural network, DMPNN). It should be noted that DMPNN is a combination of a molecule graph and ffNN, therefore all the combinations of a DMPNN and a fingerprint involve also a molecule graph.

\begin{figure}[p]
    \centering
    \includegraphics[width=0.99\textwidth]{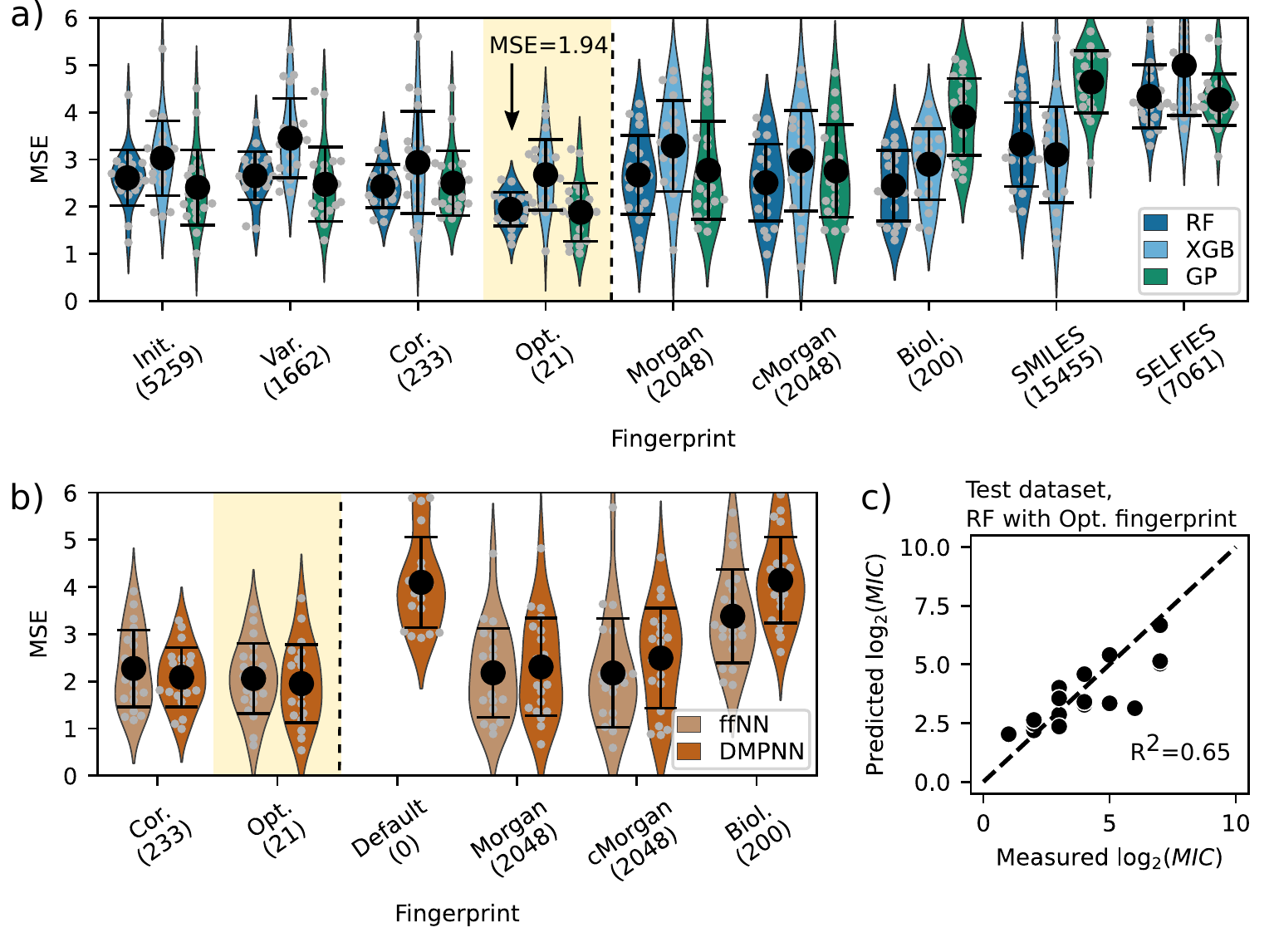}
    \caption{Comparison of different combinations of molecular representation and machine learning models, to predict antimicrobial activity of COEs. Mean-square error (MSE) for a) random forest (RF), gradient boosting (XGB), and Gaussian process (GP) regression models, as well as b) feed forward neural network (ffNN) and directed message passing neural network (DMPNN) models trained with each fingerprint option. Fingerprints (lengths in parentheses) and models are defined in the main text. MSE values are computed from 20 stratified subsampling repeats of the training dataset. Violins represent the distributions of the subsampling results, mean and standard deviation of MSE are shown in black, and individual subsample results in gray. The high-MSE tails of the distributions are clipped on MSE axes to highlight differences between the mean values. Full data are shown in ESI Figure S5. c) Test dataset prediction with the highest-performing model, RF with Opt. fingerprint, trained with the whole training dataset (test dataset withheld from training).}
    \label{fig:3}
\end{figure}

The ML models combined with the downselection fingerprints described above show improved predictive accuracies with the progressing stages of downselection (fingerprints Init. to Opt. in Figure \ref{fig:3}a, and fingerprints Cor. and Opt. in Figure \ref{fig:3}b) with the only exception of GP. The consistent improvement suggests that optimizing a fingerprint with an easily tunable model, here RF, and shifting to the final model of choice only after that, is a viable strategy to reduce model training difficulty. This is important because increasingly complex ML models, for example a range of deep-learning models, have been utilized in molecular property predictions lately. While they have provided convincing predictive accuracies, their performance may depend heavily on hyperparameter tuning. Thus, comparing and training them may be challenging \cite{Maziarka2020MoleculeTransformer, Chuang2020}, in which cases using a different model for optimizing fingerprint is convenient. Init. and Var. fingerprints have been applied only to tree-based and kernel models (Figure \ref{fig:3}a) and not to neural network models (Figure \ref{fig:3}b) because of their length. It is unlikely that neural-network-based models with very long fingerprints would have been able to fully converge into proper neuron weights with the amount of COE training data available, and therefore the results would not have been quantitatively representative of how much the fingerprint had improved.

We compared ML models with multiple molecular fingerprints used in literature: Morgan and Morgan count (cMorgan) fingerprints are extended connectivity fingerprints \cite{Morgan1965TheService} that have been widely applied for small molecule property prediction \cite{Capecchi2020OneMetabolome}. Fingerprint we denote as Biol. has been developed by Yang \textit{et al.} \cite{Yang2019AnalyzingPrediction} and utilized for biological targets, including successful application to antimicrobial activity prediction of mainly small molecules by Stokes \textit{et al.} \cite{Stokes2020}. SMILES and SELFIES fingerprints are one-hot-encoded molecule string representations of SMILES and SELFIES strings, respectively. These 2D fingerprints were tested also together with a molecular graph representation using DMPNN models. Finally, DMPNNs were in Figure \ref{fig:3}b also trained using solely a 2D molecular graph representation without additional fingerprint (Default) as illustrated in Figure \ref{fig:1}. The molecular graph involves by default 10 different basic atomic and bond type descriptors that are utilized in the message-passing scheme of the DMPNN operation. It should be noted that Figure \ref{fig:3} is not meant for universally comparing the ML model - representation combinations, because the performance depends heavily on the domain and dataset.

Figures \ref{fig:3}a-b show that the best performance for COE among all the combinations are achieved with Opt fingerprint combined with either GP or RF model. RF results in lower variance in performance and therefore it was chosen as the final model. RF with Opt. fingerprint gives mean $\mathrm{MSE}=1.94$. To give a reference to experimental units, the mean MSE transforms to RMSE of approximately 2.6\MICu{} when the order of magnitude of MIC is 2\MICu{}, or 330\MICu{} when the order of magnitude for MIC is 256\MICu{}. With the final model (trained with the full training dataset), a test dataset prediction with $\mathrm{R}^2=0.65$ is achieved (Figure \ref{fig:3}c). The resulting model is capable of singling out most of the low activity molecules that are not promising for further investigation ($\log_2(\MIC{})>=4$), which already facilitates synthesis work in a novel molecular domain. We further evaluated the ML model from the viewpoint of prospective use in predicting real-life data and as a part of an molecule optimization cycle, which are described in detail in Supplementary Materials Sections S11 and S12, respectively. We show that the ML modelling approach (1) performs well when the regressor is adapted as a classifier that aims to minimize false positives (so we don't expend experimental bandwidth on poor candidates), (2) outperforms random sampling (e.g., when we know little about a new molecular system) when utilized in a pool-based active learning loop \cite{Liang2021a} utilizing Bayesian optimization algorithm \cite{Shahriari2016a}. These results provide further reason to believe that the ML model with Opt. fingerprint is useful in the laboratory for the discovery of new antimicrobial COEs.

Figure \ref{fig:3} shows that RF performs consistently well with a range of different fingerprint options. GP provides narrowly the best mean MSE with Init. to Opt. fingerprints but results in larger variance. All the models result in lower mean MSE with Opt. fingerprint compared to other fingerprint options (test dataset predictions shown in Supplementary Figure S7). The performance of DMPNN and ffNN models is clearly improved by using Opt. fingerprint: the reference fingerprints result in large variations in model performance within the repeated subsampling folds, many of which result in practically non-usable DMPNN and ffNN models with $\mathrm{MSE}>4$ --- a model that would always output training dataset average \MIC{} would result in MSE in this range --- whereas with Opt. results in they perform close to RF and GP models with mean of $\mathrm{MSE}=1.95$ and $\mathrm{MSE}=2.06$, respectively, albeit with clearly higher variance. This demonstrates that with an optimum fingerprint, a neural network -based model can be viable even with a relatively small dataset. DMPNN might turn out to perform better than RF or GP in predicting COEs when the size of the dataset increases in future. DMPNN results in slightly lower MSE than ffNN even though the applied molecular graph message passing scheme is initially designed for small molecules \cite{Yang2019AnalyzingPrediction}. In the DMPNN-Opt. model trained here, messages are passed between the five nearest neighbor atoms in the molecule graph, which is a relatively large neighborhood area. DMPNN with a 2D molecular graph representation alone (Default) does not, however, seem to provide sufficient information to predict antimicrobial activity of COEs, but needs to combined with a supporting fingerprint (Opt.).

Many of the fingerprint-model combinations in Figure \ref{fig:3} result to high prediction errors. This happens either because the molecule representation is too long compared to the size of the data, resulting in an under-determined problem for the model, or because the fingerprint does not contain enough relevant information. Opt. fingerprint (alone or combined with a molecular graph) provides consistently the best result among the investigated fingerprints. Opt. is a short fingerprint optimized for COEs and includes many 3D descriptors in contrast to the reference fingerprints on the right-hand side of the dashed lines in Figure \ref{fig:3}a-b, all of which are longer and more general fingerprints containing only 2D information. Figure \ref{fig:3} shows that the choice of molecular representation drives the predictive accuracy even more than the choice of ML model, which has been suggested also by Wu \textit{et al.} \cite{Wu2018MoleculeNet:Learning}. The results highlight the importance of adapting the molecular representations for the target domain, even when the mechanisms of action governing the antibiotic activity are not thoroughly known.

\section{Conclusions}

We developed a model that predicts antimicrobial activity of conjugated oligoelectrolyte molecules and that can be utilized as an advisory tool to guide the synthesis of newly predicted COEs. This model was enabled by a framework that consists of four parts: (1) molecular fingerprint representation, (2) feature downselection, (3) ML model pairing, and (4) descriptor importance analysis. We applied this framework to a set of 136 COEs, using an automated molecular descriptor downselection process that is agnostic to the molecule domain. The resulting fingerprint consisted of 21 molecular descriptors, over 40\% of which are related to the three-dimensional shape of the molecules. This is in contrast to descriptors that would relate to molecular properties not dependent on shape, or descriptors related to lower-dimensional simplifications of molecular shape, such as molecule bonds with no length information. This is consistent with the presumed mechanism of action for COEs, which is intercalation into the bacterial membrane. Our results suggest the connection of 3D shape to the antimicrobial activity of COEs should be investigated further with detailed mechanistic experiments and accompanying modelling.

By developing a predictive model, we have demonstrated that ML can indeed aid development of antibiotics even in novel domains, namely with families of understudied candidates, where sparse information exists regarding the underlying mechanism and there is a limited availability of experimental data. We have shown that molecular representation drives predictive accuracy in both traditional and complex ML models: it needs to capture the critical information about the underlying mechanisms of activity of the antibiotics in order to achieve high predictive accuracy. Self-learning representations have provided impressive results in molecular property prediction lately. We suggest that fingerprinting remains a valid option or a complementary component to learned representations, provided that fingerprints are optimized for the molecular domain under investigation. This is especially in situations where there is limited data.

In closing, we presented a domain-agnostic framework to downselect and analyze a molecular fingerprint quickly, to describe novel classes of antibiotics. We investigated only one class of molecules, COEs, in this work, but 3D descriptors being so dominant in the fingerprint downselection suggests that representations and models capable of capturing 3D may be worth exploring also in other antibiotic candidate domains, particularly when the molecules are large and/or the mechanisms of action connect to molecular shape as is suspected here.

\section {Methods}
\label{sec:methods}

\subsection{Synthesis and characterization of conjugated oligoelectrolyte molecules}
\label{sec:synthesis}

The synthetic procedure of COEs can refer to previous literature \cite{Yan2016InfluenceOligoelectrolytes, Zhou2018InformedActivity, Limwongyut2020MolecularCells}. Generally, the alkylation steps were accomplished by Williamson ether synthesis with a carbonate base. The conjugated backbones of COEs were composed via Horner–Wadsworth–Emmons or McMurry coupling reactions. After final quaternization of the terminal alkyl halide groups with excess trimethylamine or other amines, the targeting COEs were obtained. Intermediates and COE products were purified using multiple approaches (including liquid-liquid extraction, column chromatography, precipitation, solvent removal under vacuum, etc.), and then characterized by NMR spectroscopy or mass spectrometry (see the Supporting Information for details).

\subsection{Forming the dataset}
\label{sec:coe}

MIC values for COEs were experimentally determined against \textit{E. coli} K12 (ATCC 47076) using the broth microdilution method as previously described in Ref. \cite{2003DeterminationDilution}. COE structures were converted into SMILES strings using ChemDraw v 19.0. SMILES strings were used in Avogadro 1.2.0 to generate 3D models of each COE. Molecular geometries were then optimized using the MMFF94 force field in Avogadro. The energy minimized 3D structures were exported as .cml files and uploaded into alvaDesc v1.0.20 for molecular descriptor calculations. A total of 5,305, 1D, 2D, and 3D molecular descriptors were calculated for each COE.

\subsection{Data preprocessing}

Dataset was preprocessed by dropping non-numeric descriptors. During the broth microdilution experiment, solutions of doubling concentrations are examined for bacterial growth. In cases where the end point of bacterial growth had not been met during the experiment, the \MIC{} value was estimated to be the next largest available value (\textit{e.g.}, when experiment resulted in $\MIC{}>256\MICu{}$, the value was estimated as $\MIC{}=512\MICu{}$). A $\log_2()$ transformation was performed on \MIC{} values since all the \MIC{} values were multiples of two. All the graphs in this work show $\log_2(\frac{\MIC{}}{1\MICu{}})$ values. It should be noted that due to the discrete nature of \MIC{} results, the effective uncertainty for \MIC{} measurement is typically one 2-base order of magnitude, and arbitrarily high in the high end for those molecules that have open-ended \MIC{} measurement results ($\MIC{}>X\MICu{}$). Finally, we examined the data for crude outliers by performing leave-one-out cross validation for the whole dataset using RF models with fixed hyperparameters. For each fold, we trained ten RF models with different initial random states, and evaluated the mean prediction error value. The molecule was dropped from the dataset as an outlier if the prediction error was higher than 3.5 $\log_2(\frac{\MIC{}}{1\MICu{}})$ units, indicating that \MIC{} prediction was more than ten times off. There were in total 5 such molecules in the original dataset used for ML model development, which are detailed in Supplementary Material Section S13.1. After the ML model had been developed, it was tested on new COE molecules (see Methods for the division of the dataset into subsets). Three of these molecules were additionally flagged by the outlier detection method but not excluded.

\subsection{Statistical down-selection of molecule descriptors}

We applied a multi-stage molecular descriptor filtering process to the dataset. The starting point of the filtering process was the dataset with all the numerical descriptors (Init.). We started by filtering out descriptors with variance less than 10\% of the mean value across the whole dataset (Var.). We continued by filtering out one descriptor of the descriptor pairs that had higher than 0.9 correlation with each other as well as descriptors that had lower than 0.05 correlation with $\log_2(\MIC{})$ (Cor.). All the correlations were evaluated as Spearman rank correlations, which evaluate any monotonic relationships, whether they are linear or non-linear ones. Correlation matrices for each fingerprint set are shown in Supplementary Material Figure S1.

\subsection{Division of the molecule dataset into subsets}

The full dataset is 136 molecules. It is divided into five parts: empty \MIC{}, outliers, training, test, and new data for which the model is applied to. At the time of initial model fitting, 33 of the molecules did not have measured \MIC{} available, and were excluded from the model fingerprint downselection and ML model training procedures. Molecules without \MIC{} measured were still included into the dataset collected in this work to provide a comprehensive picture of the COEs synthesized to date. Further five molecules were excluded during outlier detection (see Methods for data preprocessing). We designated 20\% of the remaining molecules with valid \MIC{} entry, totaling 20 molecules, into a test dataset that was held out of the model training pipeline. The remaining 78 molecules were included in the training dataset. For 15 of the molecules that did not initially have \MIC{} measured, the measurement was performed during the ML development. Therefore, the real-life use of the fully optimized ML model in predicting any COEs chosen by a scientist was tested with these 15 new molecules as described in Supplementary Material Section S11. The similarity of the outlier, training, test, and new datasets is evaluated using t-SNE similarities with Opt. fingerprint in Supplementary Materials Section S10.

Throughout the whole ML development phase, we utilized cross-validation with 20 random but stratified sub-sampling repeats for training and estimating the accuracy of the model-representation pairs. Each sub-sample was divided into 20\% validation set and 80\% train set. The sub-sampling stratification for RF, XGB, and GP was performed based on high/intermediate/low \MIC{} classes, and for DMPNN and ffNN models a similar balancing was implemented (both described in Supplementary Material Section S13.2).

\subsection{Machine learning models and hyperparameter optimizations for them}

RF and GP were implemented using Scikit-learn \cite{PedregosaFABIANPEDREGOSA2011}, XGB using XGBoost \cite{XGBoostDocumentation, Chen2016XGBoost:System}, and ffNN and DMPNN using Chemprop \cite{ChempropDocumentation}. Hyperparameter optimizations were performed using the training dataset on a high-performance computing server \cite{Reuther2018InteractiveAnalysis}. Hyperparameters for RF and XGB were optimized with Bayesian optimization with 20-fold cross validation RMSE as a target. A Python package BayesianOptimization \cite{GitHubProcesses} was utilized. GP surrogate model and expected improvement acquisition function were used for Bayesian optimization with 50 random initialization points, 300 iterations of optimization, and three restarts of the optimization. GP was implemented using Scikit-learn and this implementation had internal hyperparameter optimization with a Newtonian kind limited-memory BFGS algorithm \cite{Sklearn.gaussian_process.GaussianProcessRegressorDocumentation}. GP had Matern kernel and 50 restarts of the optimizer. Hyperparameters for the neural network models were optimized using internal implementation of chemprop that also relies on Bayesian optimization. 800 iterations and scaffold-balanced splits were utilized. The search spaces were expanded for the depth of the network (max. 12), and number of ffNN layers (max. 5). The optimized hyperparameter values are included in the GitHub repository linked to this work.

\subsection{Recursive feature elimination}

Recursive feature elimination (RFE) was applied on hyperparameter optimized RF with Cor. molecular fingerprint. RF was chosen as the base model for RFE because it is fast to ramp up and can be used without scaling the inputs. We applied a shallow RFE with cross-validation, \textit{i.e.}, RFE was first performed on the whole training dataset with steps of recursively eliminating one descriptor at a time until only one descriptor was remaining. After that, each stage was evaluated with 20-fold repeated stratified subsampling. The optimized molecular fingerprint (Opt.) was determined as the optimum arising from bias-variance trade-off, \textit{i.e.}, as the descriptor set that was at the center of the valley that resulted in the lowest mean RMSE.

\subsection{Molecular fingerprints}

Init., Var., Cor., and Opt. fingerprints were defined as described above. Mor., cMor., and Biol. fingerprints were computed with chemprop Python package \cite{ChempropDocumentation} that internally runs rdkit package \cite{RDKit}. SMILES were one-hot-encoded using our own implementation, and SELFIES were computed using  and SELFIES fingerprints were computed using selfies package \cite{GitHubChemistry}. Molecular fingerprints are used as such for RF and XGB, since these tree-based models do not require scaling of input features. GP, DMPNN and ffNN were scaled to zero mean, unit variance.

\subsection{Shapley additive explanations}

SHAP analysis was performed for the final RF model trained with the whole training dataset using SHAP Python package \cite{Lundberg2017APredictions, Lundberg2020FromTrees, GitHubModel.}. The molecules analyzed were the training dataset. Prior to the analysis, it was confirmed that the molecular descriptors in the Opt. fingerprint did not have very high correlations with each other, which could distort the analysis (see Supplementary Figure S1). SHAP provides a more fine grained method for analysing model operation compared to RF feature importance ranking (see Supplementary Figure S6) because the direction of the effects can be recognized.

\subsection{Data and code availability}

The codes for fingerprint optimization and running the comparisons between the molecule representation - ML model pairs are available in GitHub repository: https://github.com/PV-Lab/MLforCOE. Details of the molecules synthesized are listed in Supplementary Material.

\subsection{Competing interests}

The authors declare competing financial interests to include the filing of intellectual property regarding machine learning methods in general (but not the methods submitted with this paper, which are open-sourced), and start-ups concerning the use of said models and COEs.

\begin{acknowledgement}

 We thank Geraldine W. N. Chia and Chenyao Nie for performing MIC measurements. We thank Tian Xie, Allison Tam, and Regina Barzilay for providing the trained ML model from Stokes \textit{et al}. \cite{Stokes2020}, and for fruitful discussions. We thank Akshat Nigam, Daniel Flam-Shepherd, and Alán Aspuru-Guzik, for fruitful discussions regarding path message passing neural network models. We thank Gemma E. Moran for fruitful discussions and providing minimum working example code related to DMPNN modifications. The authors acknowledge the MIT SuperCloud and Lincoln Laboratory Supercomputing Center for providing HPC resources that have contributed to the research results reported within this paper. Q.L. acknowledges generous funding from TOTAL S.A. for supporting his graduate research. Z. R. is supported by the National Research Foundation, Prime Minister’s Office, Singapore under its Campus for Research Excellence and Technological Enterprise (CREATE) program through the Singapore Massachusetts Institute of Technology (MIT) Alliance for Research and Technology’s Low Energy Electronic Systems research program. A.T., N.H, Z.L., S.S., and T.B. acknowledge support from DARPA under Contract No. HR001118C0036, and TOTAL S.A. S.J. acknowledge funding from the Accelerated Materials Development for Manufacturing Program at A*STAR via the AME Programmatic Fund by the Agency for Science, Technology and Research under Grant No. A1898b0043. Work at NUS was supported through a Startup Grant to G.C.B. (R-143-000-A97-133). Work at UCSB was supported by the Institute for Collaborative Biotechnologies through grant W911NF-09-D-0001.

\end{acknowledgement}

\begin{suppinfo}

Additional experimental details, materials, and methods, including information on the descriptors at each downselection stage, test dataset predictions of the fitted models, detailed description of outlier selection and subsampling processes, and experimentally measured minimum inhibitory concentrations as well as nuclear magnetic resonance or mass spectrometry spectra of the synthesized molecules.

\end{suppinfo}

\clearpage

\bibliography{references_manually_modified_20211006.bib}{}

\clearpage


\end{document}